\documentclass[twocolumn]{aastex61}
\usepackage{graphics,graphicx}
\usepackage{epsfig}
\usepackage{amsmath}
\usepackage{rotating}
\usepackage{array}
\usepackage[caption=false]{subfig}
\usepackage{geometry}
\usepackage{natbib}
\usepackage[english]{babel}
\usepackage{appendix}
\usepackage{soul}

\received{\today}
\revised{\today}
\accepted{\today}
\submitjournal{ApJ}
\shorttitle{self-calibration of solar interferometry data}
\shortauthors{Mondal et al.}

\newcolumntype{C}[1]{>{\centering\let\newline\\\arraybackslash\hspace{0pt}}m{#1}}

\newcommand\plotfivecropped[5]{%
	\centering 
	\leavevmode 
	\includegraphics[trim={0 3cm 0.7cm 2.9cm},clip,scale=0.3]{#1}
	\hfil
	\includegraphics[trim={2.3cm 4.1cm 0.8cm 3cm},clip,scale=0.337]{#2}
	
	\centering
	\includegraphics[trim={2.2cm 4.6cm 0.7cm 3.1cm},clip,scale=0.36]{#3}
	 \hfil
	 \includegraphics[trim={2cm 4cm 0 3cm},clip,scale=0.37]{#4}

	   \centering
	\includegraphics[trim={0 4cm 0 3.2cm},clip,scale=0.37]{#5}
}

\newcommand\plotsixcropped[6]{%
	\centering 
	\leavevmode 
    \includegraphics[trim={0.1cm 0 1.2cm 1.2cm},clip,height=5.7cm, width=6cm]{#1}
	\hfil
	\includegraphics[trim={2.3cm 2.4cm 0.8cm 2.5cm},clip,scale=0.3]{#2}
	
	\centering
	\includegraphics[trim={2.3cm 2.4cm 0.8cm 2.5cm},clip,scale=0.3]{#3}
	 \hfil
	 \includegraphics[trim={2.3cm 2.4cm 0.8cm 2.5cm},clip,scale=0.3]{#4}
	 
	\centering
    \includegraphics[trim={2.3cm 2.4cm 0.8cm 2.5cm},clip,scale=0.3]{#5}
    \hfil
	 \includegraphics[trim={2.3cm 2.4cm 0.8cm 2.5cm},clip,scale=0.3]{#6}
}

\begin{document}

\title{Unsupervised generation of high dynamic range solar images: A novel algorithm for self-calibration of interferometry data}
\correspondingauthor{Surajit Mondal}
\email{surajit@ncra.tifr.res.in}

\author{Surajit Mondal}
\affil{National Centre for Radio Astrophysics, Tata Institute of Fundamental Research, Pune 411007, India}

\author{Atul Mohan}
\affil{National Centre for Radio Astrophysics, Tata Institute of Fundamental Research, Pune 411007, India}

\author{Divya Oberoi}
\affil{National Centre for Radio Astrophysics, Tata Institute of Fundamental Research, Pune 411007, India}

\author{John S. Morgan}
\affil{International Centre for Radio Astronomy Research, Curtin University, GPO Box U1987, Perth, WA 6845, Australia}

\author{Leonid Benkevitch}
\affil{ Massachusetts Institute of Technology- Haystack Observatory, Westford, Massachusetts, USA}

\author{Colin J. Lonsdale}
\affil{ Massachusetts Institute of Technology- Haystack Observatory, Westford, Massachusetts, USA}

\author{Meagan Crowley}
\affil{University of Massachusetts, Boston, USA}

\author{Iver H. Cairns}
\affil{University of Sydney, Sydney, Australia}

\begin{abstract}

Solar radio emission, especially at metre-wavelengths, is well known to vary over small spectral ($\lesssim$100\,kHz) and temporal ($<1$\,s) spans. It is comparatively recently, with the advent of a new generation of instruments, that it has become possible to capture data with sufficient resolution (temporal, spectral and angular) that one can begin to characterize the solar morphology simultaneously along the axes of time and frequency. This ability is naturally accompanied by an enormous increase in data volumes and computational burden, a problem which will only become more acute with the next generation of instruments such as the Square Kilometre Array (SKA). The usual approach, which requires manual guidance of the calibration process, is impractical. Here we present the ``Automated Imaging Routine for Compact Arrays for the Radio Sun (AIRCARS)", an end-to-end imaging pipeline optimized for solar imaging with arrays with a compact core. 
We have used AIRCARS so far on data from the Murchison Widefield Array (MWA) Phase-I. The dynamic range of the images is routinely from a few hundred to a few thousand. In the few cases, where we have pushed AIRCARS to its limits, the dynamic range can go as high as $\sim$75000. The images made represent a substantial improvement in the state-of-the-art in terms of imaging fidelity and dynamic range. This has the potential to transform the multi-petabyte MWA solar archive from raw visibilities into science-ready images. AIRCARS can also be tuned to upcoming telescopes like the SKA, making it a very useful tool for the heliophysics community.
\end{abstract}

\section{Introduction}

Imaging the radio Sun with good fidelity is an intrinsically hard problem.
It has structures spanning a large range of angular scales; and the intrinsic brightness temperatures ($T_B$) associated with different emission mechanisms also span many orders of magnitude.
From studies using the solar dynamic spectrum, it has long been known that the coronal emission can have structures with fractional bandwidths\footnote{$\delta \nu/\nu_0$, where $\nu_0$ refers to the central frequency of observation and $\delta \nu$ to the observing bandwidth.} of the order of a percent and sub-second time scales.
Recent studies \citep[e.g.][]{sharma18,suresh17} have shown the presence of such structures at lower flux densities than were observed before.
Tracing these small scale changes in solar radio images necessarily requires high fidelity imaging with sufficiently high time and frequency resolutions.
Additionally, the Sun is much brighter than the usual calibrator sources, which leads to complexities in achieving robust calibration.
These challenges, further compounded by the limitations imposed by the earlier instruments, have meant that, with few exceptions, radio imaging studies of the active Sun have been limited to studying the brightest sources of emissions at a few discrete frequencies \citep[numerous examples in][]{pick08}, and those of the quiet Sun have usually required integration times of many hours \citep[e.g.][]{mercier2009}.

Riding on the wave of enormous advances in digital signal processing and increased affordability of computing, many new metric radio interferometers such as the LOw Frequency ARray \citep[LOFAR;][]{harlem13} operating in the ranges 10-80 and 120-240\,MHz, the Long Wavelength Array \citep[LWA;][]{kassim2010} in the 10-88\,MHz band and the Murchison Widefield Array \citep[MWA;][]{lonsdale09,tingay2013} in the 80-300\,MHz band have become available comparatively recently. 
They represent significant improvements in instrumental capabilities and are well suited for solar imaging work with high temporal and spectral resolution at low radio frequencies.
These instruments are already producing interesting scientific results \citep[e.g.][]{morosan14,morosan15,beltran15,morosan16, kontar17, morosan17, mccauley17, suresh17, cairns18, sharma18, mohan18, zucca18}.
{ The advent of Atacama Large millimeter and sub-millimeter Array, with its large number of elements, is leading to a large improvement in the quality of solar snapshot images at mm wavelengths \citep[e.g.][]{nindos18}.}

Of all the new generation instruments, the MWA design is best suited for high fidelity spectroscopic snapshot solar imaging.
However, these information rich data are necessarily much more voluminous and pose a large computational burden. 
The raw MWA data over a bandwidth of 30.72\,MHz and a duration of 1 minute at time and frequency resolution of 0.5\,s and 40\,kHz is about 28 GB and can be used to make about $10^5$ images.
Manual analysis, which has been the norm in traditional radio interferometry, is impractical in such cases.
A detailed science exploration of these data requires automated analysis tools.
Standard interferometric tools have been adapted to MWA solar observations with considerable successs \citep[e.g.][]{mccauley17,rahman18}.
Our experience has been that by tuning the analysis to the needs of solar imaging, it is often possible to significantly improve the quality of images obtained and also the likelihood of successfully imaging any given data set. 
This forms the motivation for the work presented here.
In addition, this work marks our first step in reducing the effort involved in going from raw radio interferometric data to the corresponding science ready imaging data products. 
We believe that the very limited availability of such data products have contributed to the infrequent usage of low frequency radio interferometric imaging data in multi-wavelength solar studies, even though the scientific merits of these data are well established \citep[e.g.][etc.]{bastian01,pick08,mercier15}.
The algorithm presented here is general and can be used for any array with a central dense core, such as the future SKA-Low, which includes solar and heliophysics as its primary science targets \citep[][]{nakariakov15,nindos18}.

In addition, in order to reduce the tedium involved, interactive analysis tends to make some assumptions which can easily be relaxed by an automated pipeline (Sec. \ref{pipeline}), leading to significant improvements in the imaging dynamic range (DR) obtained.
These high DR images are essential for achieving some of the science targets of interest \citep[e.g.][]{bowman13,nindos18}.

This paper is organized as follows: Section 2 and Section 3 describes the algorithm and the pipeline respectively. Section 4 discusses the data used for testing this pipeline and showcases the high DR images obtained using it. In Section 5, we comment on the various issues in calibrating the data. Section 6 summarizes and concludes this work.





%

\section{Algorithm description}\label{algorithm}

We have developed an algorithm to calibrate the interferometric cross-correlations (visibilities) accurately to yield high DR radio images.
There are two steps in the calibration procedure. The first step, namely \textit{Amplitude Decorrelation Correction}, is required because of instrument specific requirements of the MWA. This algorithm does a baseline based correction and is described in section \ref{sec:decor}. Section \ref{sec:selfcal} describes the algorithm for doing an antenna-based gain calibration in an automated manner.

\subsection{Amplitude Decorrelation Correction} \label{sec:decor}

Radio interferometer correlators typically correct for geometric delays and delays due to varying cable lengths from the antenna to the receiver so that the electromagnetic waves received from a particular point on the sky (the `phase center') are combined coherently.
The MWA correlator \citep{ord15} does not have a delay model and simply cross correlates the signal received from each antenna pair.
The array is phased up towards the true pointing direction\footnote{or any other chosen coordinates such as those of the Sun} of the tiles\footnote{{Each element of the MWA comprises of a $4 \times 4$ array of dual polarization dipoles, referred to as `tiles'}} during offline processing by suitably rotating the phases of the visibilities \citep[][]{offringa15}.
This operation is equivalent to adjusting the delays pre-correlation \citep{morgan11} except for decorrelation losses due to finite channel width and integration time.
However, as the correction is applied after the calculation of visibilities, it can only be done with the granularity of the time and frequency resolution post correlation\footnote{these have a maximum frequency resolution of 10\,kHz, a maximum time resolution of 0.5s, and a minimum product of the two of 20\,000.}, and cannot correct for the phase variation across one spectral channel.
This leads to a reduced correlation amplitude, $a_{\mathrm{cor}}$, which is given by
\begin{equation}
  \label{eq:corr_factor}
  a_{\mathrm{cor}} = \frac{1}{\Delta \nu}\int_{\Delta \nu/2}^{\Delta \nu/2}e^{i 2 \pi\nu\tau} \mathrm{d} \nu = \mathrm{sinc}\left(\tau \Delta \nu\right).
\end{equation}
Here $\tau = \tau_2 - \tau_1 + \tau_{w_{12}}$, $\tau_{1,2}$ are the cable delays for the antennas $1$ and $2$; $\tau_{w_{12}}$ is the delay due to $w$ component of the baseline comprising of antenna $1$ and antenna $2$ and $\Delta \nu$ is $40\,kHz$, the bandwidth of the frequency channel.
Equation \ref{eq:corr_factor} assumes a rectangular bandpass filter, which is justified in the case of the MWA, which uses digital polyphase filters.

Decorrelation due to a change in the visibility phase across a channel (or across a time integration) is well known in radio astronomy, since it arises when there is a source at some distance from the phase centre \citep[see e.g.][]{bridle99}.
In this case, the decorrelation is related in a straightforward way to baseline length.
In the image plane the decorrelation effect mimics the source being extended, hence this effect is commonly referred to as ``smearing''.
Since identical baselines will suffer exactly the same decorrelation, the effect on the achievable dynamic range in snapshot images will be minimal.
This is also true of the time-average smearing of our data, which in any case is at least an order of magnitude smaller than the decorrelation losses due to channelisation.

\begin{figure}
    \centering
    \includegraphics[width=1.1\columnwidth]{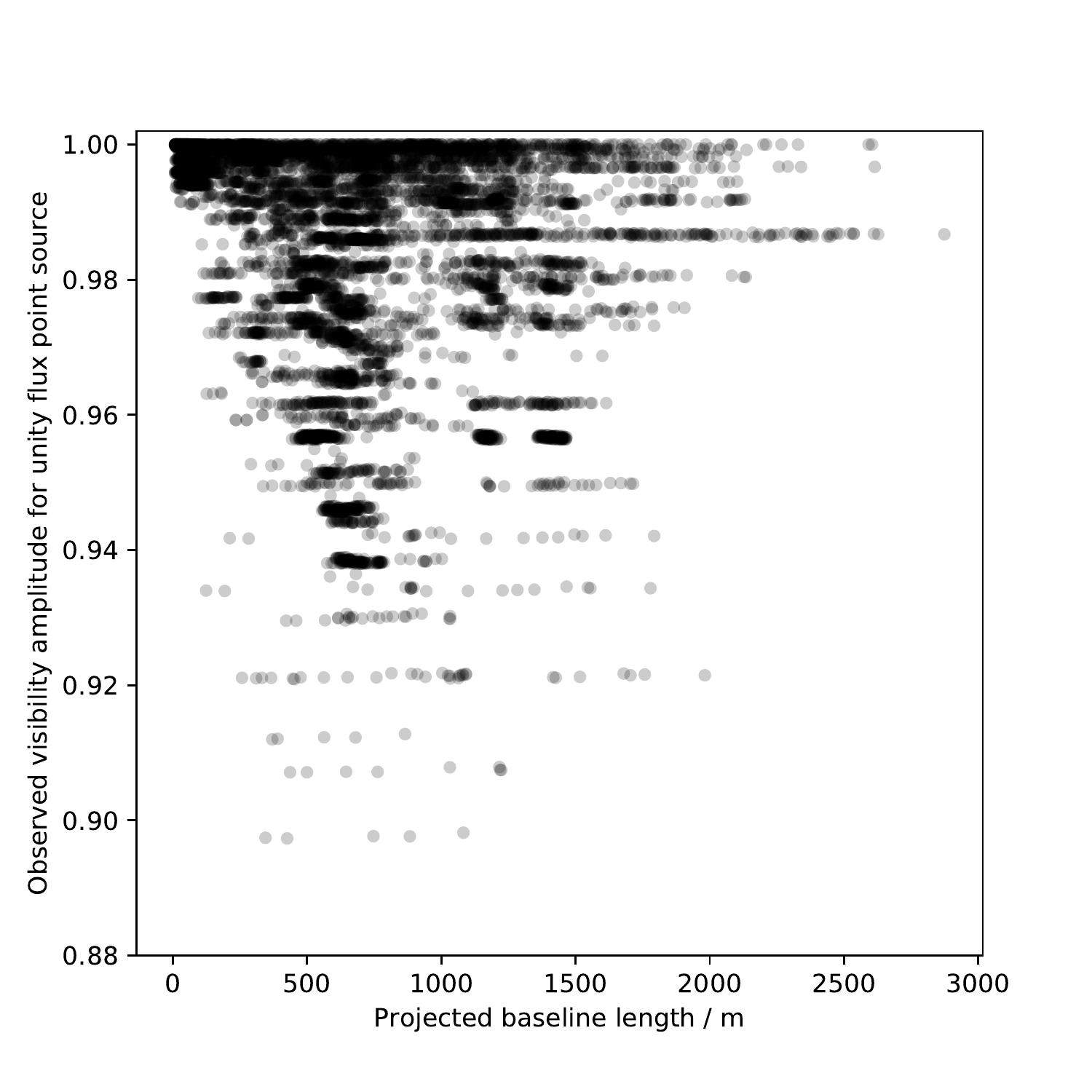}
    \caption{
    { Distribution of $a_{\mathrm{cor}}$ as a function of the projected baseline length as seen from the direction of the source. 
    The lack of a clear relationship between $a_{\mathrm{cor}}$ and the length of the projected baseline is self evident.
    }}
    \label{fig:acor}
\end{figure}

The distribution of $a_{\mathrm{cor}}$ for a typical solar observation is shown in Fig. \ref{fig:acor} with minimum value of $a_{\mathrm{cor}} \approx 0.9$ corresponding to a 10\% decrease from the maximum value of 1.0.
It is immediately clear that the decorrelation is not a straightforward function of baseline length, and two identical baselines could have very different decorrelation losses. 
Thus it is imperative to correct these decorrelation losses if high dynamic range is to be achieved.
Although simple, this is not universally applied to MWA since it can only be correct for one point on the sky.
Correcting these issues across the full field of view is not straightforward, although it has been attempted for MWA data \citep[ch. 6]{line17}.
Fortunately for solar imaging, the correction does not change significantly across the effectively small field of view and so the corrected visibility $V_{\mathrm{cor}}$ is simply 
\begin{equation}
  \label{eq:visibility_corrected}
  V_{\mathrm{cor}} = \frac{V_{obs}}{a_{\mathrm{cor}}}, 
\end{equation}
and this has been shown by \citet{deller11} to perfectly reproduce the visibility, the only penalty being a modest loss in signal to noise.
The high flux density of the Sun also ensures that the flux contribution from all other sources can effectively be ignored.
In our experiment, we saw an improvement of 30\% in the image dynamic range after correction for this decorrelation.
A similar level of improvement is expected irrespective of the details of the solar emission morphology as $a_{cor}$ depends only on observing geometry and cable lengths.


\subsection{Calibration of antenna gains} \label{sec:selfcal}

The basic objective of calibration is to determine and remove the instrument specific artefacts, so as to isolate the true sky signal. 
All of the instrumental effects can be captured in a single complex number per antenna, referred to as antenna gain $g_i$, where $i$ is an index referring to the $i^{th}$ antenna.
As interferometric measurements are differential in nature, $g_i$ is measured with respect to a reference antenna.
Propagation through the terrestrial atmosphere also modifies the incident radiation. 
At the low radio frequencies of interest here, it is the ionosphere which plays the dominant role. 
From a calibration perspective, the propagation effects are indistinguishable from instrumental ones, and hence are grouped together. 
We refer to the instrumental part of $g_i$ as $g_{i,inst}$ and the propagation part as $g_{i,prop}$.
$g_i$ is then defined as $g_i=g_{i,inst} \times g_{i,prop}$.

Conventionally, the antenna gains can be inferred using the signals received from one of several astronomical ``calibrator'' sources, whose positions, structures and flux densities are well known.
As the $g_i$ measured using calibrators come from different, though nearby, times and directions, there remain some differences between the $g_{i}$ thus derived and those towards the target source.
A method called self-calibration is traditionally used to overcome this limitation.
Unlike usual calibration, self-calibration treats both the $g_i$ and the sky model (i.e. the brightness distribution of the target source) as free parameters, while iteratively minimizing the differences between the observed data and the sky model.
The convergence of this method, therefore relies crucially on the quality of the initial sky model.

\subsubsection{Considerations specific to solar calibration}
Solar observations depart from these norms in a few significant ways.
The Sun is such a strong source that the calibrators must be observed at sufficiently large angular distances so that the flux density picked up by the primary beam side-lobes towards the direction of the Sun is weak enough to not contaminate the calibrator measurements.
For a wide FOV instrument with large primary beam side-lobes, like the MWA, this usually implies that the calibrators must be observed before sunrise or after sunset. 
In addition, the solar flux density is so large that few instruments have signal paths with linear ranges large enough to observe typical calibrator sources and the Sun with identical gain settings.
Hence, to use the MWA signal chain for solar observations, the solar signal is sufficiently attenuated so that the signal strength roughly matches that of typical night time sky observation when it is presented to the receiver.
We determine the flux scale from independent measurements using a non-imaging technique described in \citet{oberoi17}.
The methodology used to transfer these flux densities to solar images is described in \citet{mohan17}, along with the procedure to make brightness temperature ($T_B$) maps.
The calibration and imaging algorithm presented here represents a large improvement over our earlier works.

\subsubsection{Self-calibration}\label{sec:self_calib}
In the usual calibration procedure, where calibrator sources are used, the sky model is well known.
The antenna gains are obtained by minimizing the { quantity $\phi$ which is given by}
\begin{equation}
    \phi=\sum_{pq}\left |V_{pq}-g_pg_qV_{pq}^{M}\right |^2 \label{eq:minimizer}.
\end{equation}
$V_{pq}$ and $V_{pq}^M$ are the visibility observed and the model visibility for baseline $pq$ respectively.
In absence of an accurate sky model, both sky model and the antenna gains in Equation \ref{eq:minimizer} can be  treated as free parameters and can be constrained by { minimizing $\phi$}.
Equation \ref{eq:minimizer} can be rewritten in the following form where both $g_i$ and sky model are free parameters:
\begin{equation}
    \phi=\sum_{pq}\left |V_{pq}-g_pg_q F I^{M}\right |^2 \label{eq:selfcal_minimizer}.
\end{equation}
F is the Fourier transform operator and $I^M$ is the model sky.
A basic assumption made while minimizing Equation
\ref{eq:selfcal_minimizer}  is that the 
sky emission can be described well by a sufficiently small number of degrees of freedom.
As the number of free parameters increases, and eventually become comparable to the number of constraints, or the measured visibilities, the self-calibration becomes increasingly poorly constrained. 
In practice, this approximation holds and the number of measured visibilities is generally much larger than the number of free parameters. 
Then an iterative process is followed to { minimize $\phi$ using Equation \ref{eq:selfcal_minimizer}}.
The steps of the iterative process are:

\begin{enumerate}
   \item Apply the best available calibration to the visibilities and do a Fourier inversion to generate an image of the sky.
   \item Use this image to build a model for the sky, typically using a CLEAN based deconvolution algorithm.
   \item Evaluate if the rate of improvement in image quality has slowed down sufficiently for this process to be deemed converged. If not, proceed to the next step.
   \item Use this sky model to find improved values of $g_i$ by  { minimizing $\phi$ using  Equation \ref{eq:minimizer}}.
    \item Go back to the first step.
\end{enumerate}

This iterative process is commonly known as self-calibration \citep[][]{pearson1984} and is routinely used in radio interferometry. 
Equation \ref{eq:selfcal_minimizer} is highly non-linear and non-convex.
Hence there is always a possibility of self-calibration converging to a local minimum, implying physically that the derived source model is not accurate.
Hence it is essential to have a reliable estimate of the sky model before the start of the self-calibration cycle.

\subsubsection{Taking advantage of the dense array cores} \label{sec:dense_core}
This calibration algorithm developed here takes advantage of the fact that for arrays with dense cores, the $g_{i,prop}$ of the large number of nearby elements are likely to be very similar to each other. 
Thus, if the reference antenna is chosen from this dense core, the numerical values of $g_{i,prop}$ for other core antennas is very close to unity.
For arrays like the MWA, which have stable $g_{i,inst}$, calibrator observations with large separations in time and direction can be used to calibrate the array.
Since the $g_{i,prop}$ of the antennas in the core is close to unity and the resolution offered by the core alone is very coarse, the calibrated data from this part of the array can be used to make a reliable and simple source model.
For antennas further from the reference antenna, the line of sight (LOS) to the target source becomes increasingly separated from that of the target antenna, and the $g_{i,prop}$ will increasingly diverge from unity.
The $g_i$ determined from the self-calibration of the antennas in the core provide a robust starting point for a bootstrap approach to include longer baselines.

\begin{figure}
    \centering
    \includegraphics[scale=0.38]{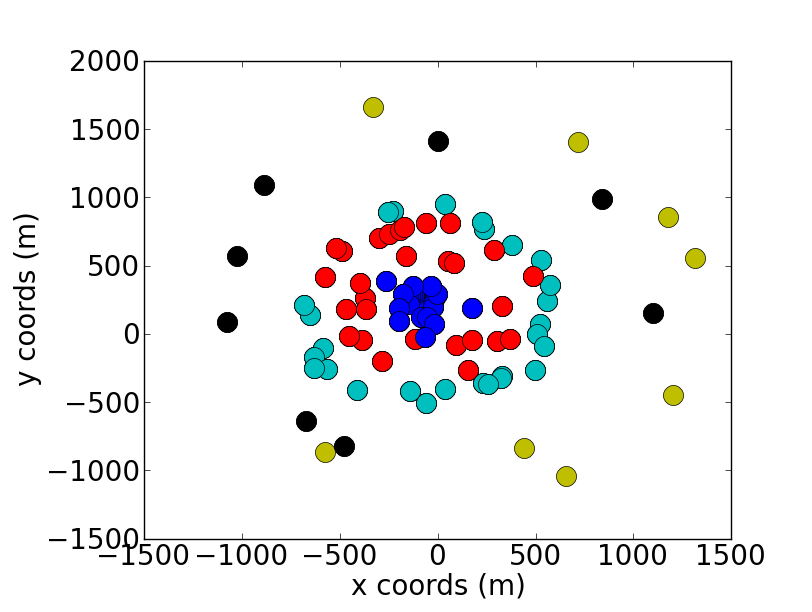}
    \caption{The location of the antennas progressively added in subsequent steps {of self-calibration} for MWA Phase-I. The antennas marked by blue are used in the first step, and antennas at increasingly large distances from a fiducial array centre are added in subsequent steps. Antennas marked by red are added in the next step, followed by those shown in cyan, black and yellow.}
    \label{fig:antenna_addition}
\end{figure}

All of the antennas are always used when solving for the antenna gains.
However, the source model is only constructed using an iteratively increasing number of antennas.
The iterative process of including antennas at increasing separation from the reference antenna and self-calibration, is repeated until the entire array is used for making the source model.
The progression of adding more antenna that we have adopted for the case of MWA Phase-I is illustrated in Fig. \ref{fig:antenna_addition}.
For each iteration in this process, the source model is built afresh, after correcting the data using the $g_i$ derived in the previous iteration (s).
At the end of this process we have a reasonable calibration for all antennas.
We use this as a starting point for the usual self-calibration cycle, which includes all the antennas to begin with.

In practise, we find that the algorithm converges to stable $g_i$ and reasonable looking images even without using calibrator observations in all the cases we have tried.
However, the use of calibrator observations is essential in order to retain astrometric information in the final images.
Using the $g_i$ derived with a calibrator observations is also expected to lead to a faster convergence of the self-calibration cycle.

In spite of the computational burden it imposes, we find it useful to iteratively add a small number of uncalibrated antennas to a larger set of calibrated antennas as we proceed in our self-calibration cycles. 
We find that with this approach, the self-calibration process always converges and the high imaging DR obtained assures us that it has converged to good solutions.
Once such solutions have been obtained for a given timeslice, we use them for initial calibration of nearby timeslices.
The first timeslice to be calibrated is referred to as $t_{ref}$ henceforth. 
The evolution in gains over the few minutes of observing duration are generally small and can be easily corrected in a few self-calibration iterations.
If one has reason to believe that the gains will not change by more than the desired level of accuracy between two timeslices of interest, one can choose not to do self-calibration and apply gains from a nearby timeslice.
The maximum time difference between which antenna gains are not expected to change is referred to as $\tau_{cal}$.

One needs to define a convergence criterion for when to stop the self-calibration process.
For this, we have chosen a criterion based on the rate of improvement of the imaging DR.
Imaging DR is defined to be the ratio of the intensity of the brightest point in the map to the rms in the image, $\sigma$, far away from the Sun.
Self-calibration is considered to have converged if the DR of a few previous iterations have changed by less than a user-specified amount, when compared to that for the present iteration.

\begin{figure*}
\centering
\includegraphics[scale=0.25]{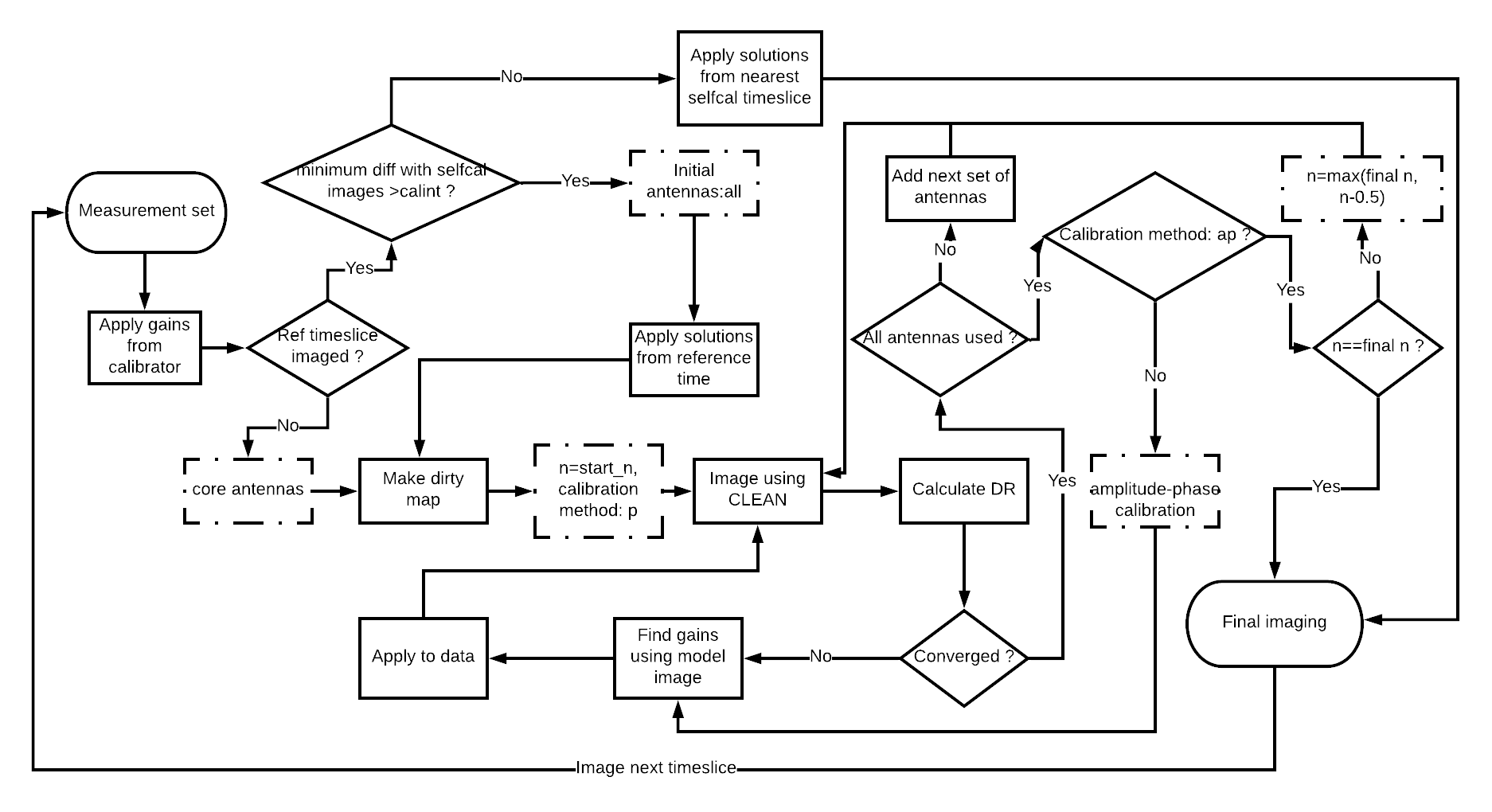}
		\caption{Broad work-flow in the calibration module. The rhombus shaped boxes denote a decision making step. A process is denoted by rectangles with solid line. Steps where a parameter value is set is denoted by a rectangles with broken lines.
		In this figure, we denote $\tau_{cal}$ by \textit{calint}, $n_{start}$ by \textit{start\_n} and $n_{final}$ by \textit{final\_n}.}
\label{fig:calibration_module}
\end{figure*}

An automated calibration algorithm needs the ability to terminate a diverging or non-converging self-calibration run.
Self-calibration usually diverges when spurious features creep into the source model.
In our experience, this is usually accompanied by a large subsequent decrease in the DR.
The DR of images is hence monitored and if a large and persistent decrease in DR is found, self-calibration is stopped.
In addition, self-calibration is also stopped if the number of iterations exceeds the maximum number of iterations allowed by the user.
Although in our experience with MWA Phase-I data, we have never encountered such a situation, we have included this as a fail-safe mechanism.
Under either of these circumstances, we re-initiate the self-calibration process using the original approach of iterative addition of antennas.
If this too does not converge, the timeslice in question is recorded in a logfile for human inspection.
A flowchart of the calibration module is given in Fig. \ref{fig:calibration_module}.

\section{Automated Imaging Routine for Compact Arrays of the Radio Sun}\label{pipeline}

We have implemented the above calibration algorithm, followed by the imaging step, in a pipeline which we have named ``Automated Imaging Routine for Compact Arrays of the Radio Sun" (AIRCARS), which can run completely unassisted.
AIRCARS is written in Python and the bulk of the radio interferometry specific functionality it uses comes from the Common Astronomical Software Analysis \citep[CASA,][]{mcmullin07}. 
Although it uses functionality provided by CASA, its architecture offers the user flexibility to use other radio interferometric software packages.
AIRCARS also implements an approach to parallelisation to allow the users to make optimal use of modern multi-core machines.

AIRCARS is an easy-to-install software and can run on any machine and operating system which supports CASA.
It follows a modular architecture, making it easy to modify, adapt, build on and maintain.
It aims to hide much of the complexity of the details of interferometric imaging from someone with little radio experience, while trying to strike the balance to provide enough control to those familiar with intricacies of interferometric imaging.
A brief description of some of its key parameters follows.
Although the algorithm implemented in AIRCARS (and also described in Section \ref{sec:selfcal}) is general, the present implementation has been tuned for the MWA.

\begin{itemize}
    \item{$n_{start}$: The CLEAN threshold is specified as $n$ times the rms of image in previous self-calibration iteration. This parameter specifies the starting value of $n$ at the beginning of the self-calibration process. Based on our experience with the MWA data, it is set to a default value of 10.}
    
\item{$n_{final}$: Self-calibration is said to be over when $n$ reaches this value.
This parameter needs to be chosen with care.
Too low a value of this parameter increases the possibility of including noise in the source model; on the other hand, too high a value can terminate the self-calibration cycle before it has had the opportunity to include the weaker features in the model. 
The appropriate choice of $n_{final}$ depends on the science objective, and also on the nature of solar emission likely to be present in the data.
Our choice of the default value has been optimized for large scale imaging exercises.
This information resides in a table which lists the values of the maximum pixel in the dirty maps and the suggested value of $n_{final}$ as ordered pairs.
The appropriate value is chosen from this table by linear interpolation across the neighboring ($\log_{10}$(maximum pixel), $n_{final}$) values.
This table is a part of AIRCARS and has been built based on our experience with MWA Phase-I data.
AIRCARS provides the users the flexibility to supply their own choices for these values by providing a table in an identical format.}

\item{$\tau_{cal}$: This defines the duration (in seconds) within which, rather than computing fresh self-calibration solutions, those available from the nearest time are applied before the data is passed on to the Imaging Module (Section \ref{imaging}).
Ideally, this should correspond to the duration over which the antenna gains are expected to remain stationary. 
The default value of this parameter is 10\,s.}

\item{\textit{limit\_dyn}: Once this DR is achieved, the self-calibration process is terminated and
the calibrated data are used to generate the final image.}
\end{itemize}

The choice of the step size in which the antennas are added depends on their distribution in the array and the user requirements.
To help with this we include a code in AIRCARS to compute the histogram of the antenna separations from the median location. Once the user chooses the optimum number of antenna addition steps and verifies the antenna distribution in each step by inspecting the histogram, the results are written to a file, which can be read by AIRCARS.
These files for MWA Phase-I and MWA Phase-II high resolution array are already provided in AIRCARS.

\subsection{Imaging Module}\label{imaging}
The imaging module uses the CASA task CLEAN to generate the source model.
It is a completely customisable module which provides full control of imaging choices to the user, ranging from the weighting schemes and cleaning threshold to the choice of using w-projection. 
{
The present implementation of this module is limited to generating Stokes I maps.
This limitation comes not from the algorithm itself, but from lack of a suitable implementation of polarization calibration.
For an array comprising element like the MWA tile, polarization calibration can be a fairly involved.
In addition to the usual relative amplitude and phase calibration of the two linear polarizations, one also needs to correct for the potentially substantial polarization leakages arising from the geometry of the wide field of view dipole array and from the mutual coupling between the elements of the array of dipoles.}

{The MWA project has made considerable progress in building a detailed understanding of the tile beams, including their polarization leakage properties \citep[e.g.][]{neben15, line18}, and arrived at models for these beams for polarimetric calibration and imaging \citep[e.g.][]{sokolowski17, lenc17, lenc18}. 
Efforts are underway to incorporate polarization calibration algorithms along similar lines in AIRCARS.}

\subsection{Making AIRCARS robust} \label{sec:conservative}

An unsupervised pipeline does not have the opportunity to benefit from the wisdom of an expert human for identifying where in the processing something is beginning to go wrong and recover from it by modifying the analysis appropriately. In order to make such an implementation robust it is essential to make what might be considered conservative choices in other contexts. This essentially leads to including additional steps in the processing which may not be needed for `well behaved data' or during a supervised analysis. This is especially true for self-calibration cycles which cannot recover from the presence of spurious features in the model. We lean towards reliability and robustness, even though it comes at a larger computational cost. These choices include:
\begin{enumerate}
    \item Using stronger convergence criteria when calibrating the $t_{ref}$ slice. 
    \item Using Hogbom CLEAN \citep[][]{hogbom1974} rather than the computationally leaner Clark CLEAN \citep[][]{clark1980}, to ensure that the point spread function (PSF) is  subtracted from the entire image \citep{cornwell99}.
    \item Reducing the CLEAN loop-gain from its usual value of 0.1 to 0.05, to reduce the possibility of picking up spurious sources in complicated source models.
    \item Performing frequent CLEAN major cycles to limit the errors incurred.
    \item Using natural weighting during self-calibration to make the optimal use of the sensitivity of the data \citep{cornwell99} .
    \item Use of multi-scale algorithm in CLEAN to represent features spanning large range of angular scales \citep{cornwell08}.
\end{enumerate}

\section{Results}\label{sec:results}

\begin{figure*}[ht]
	\plotfivecropped{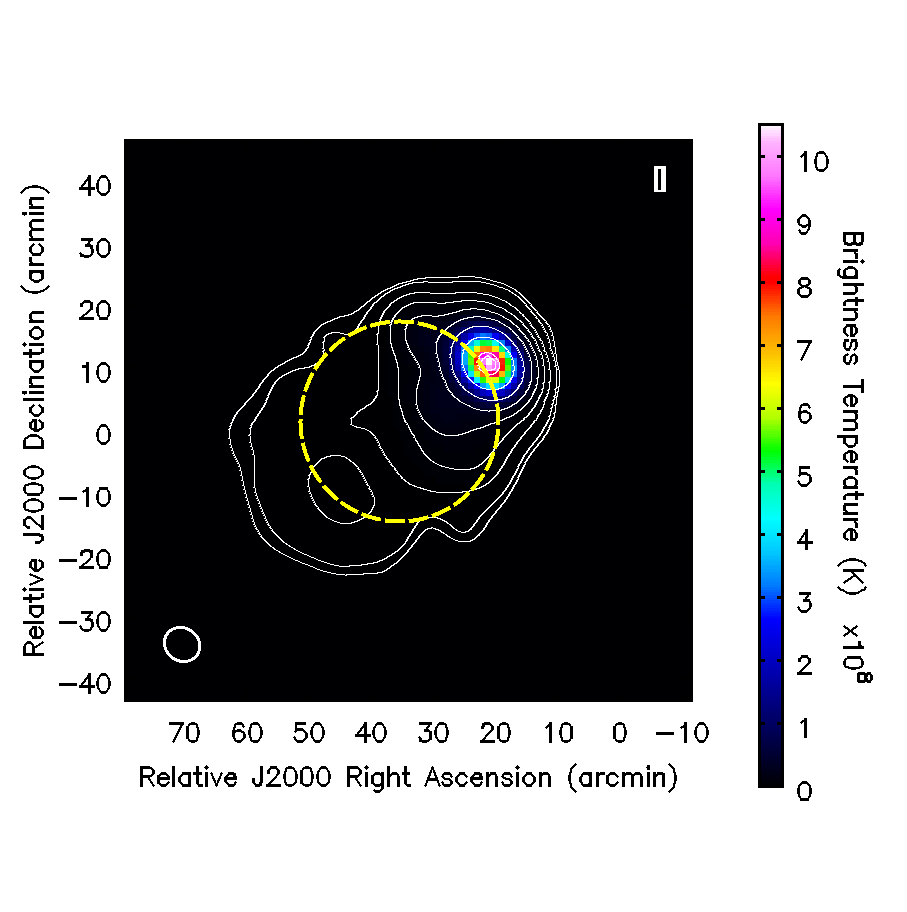}{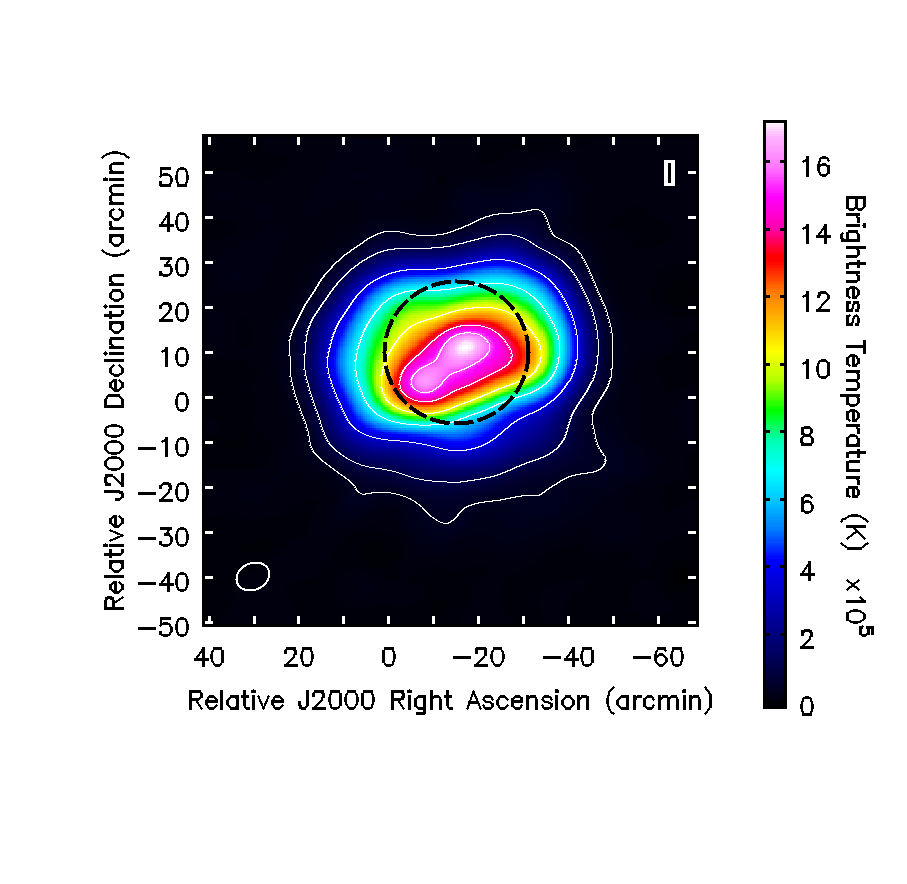}{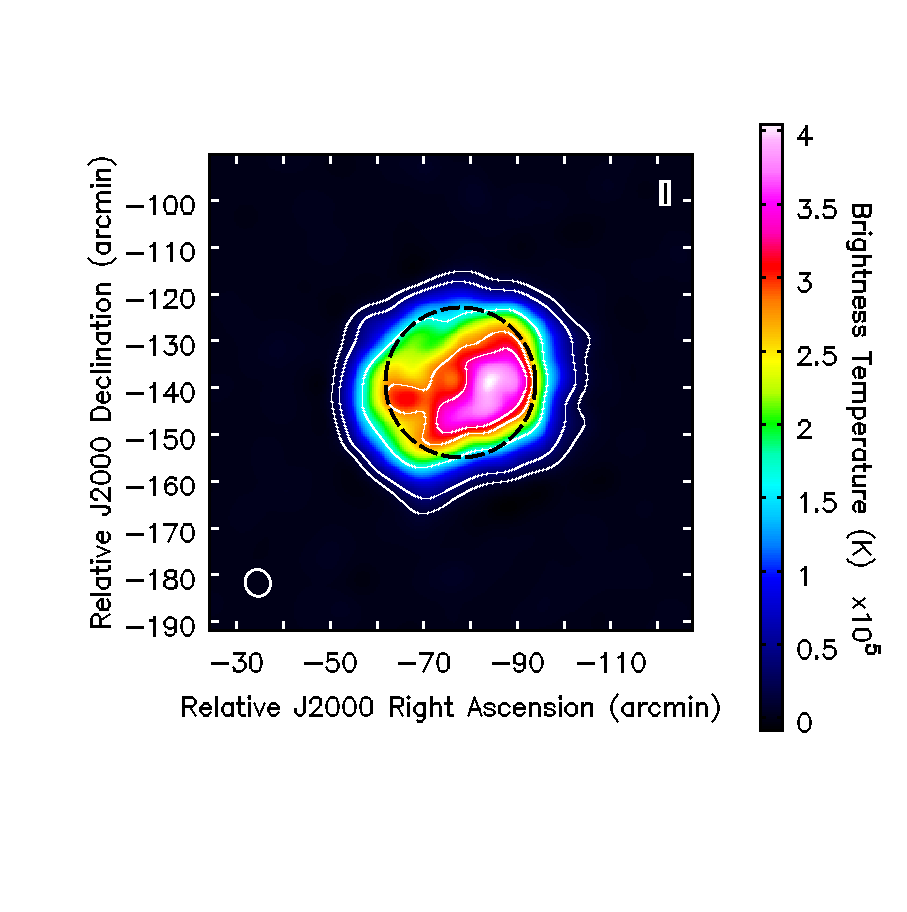}{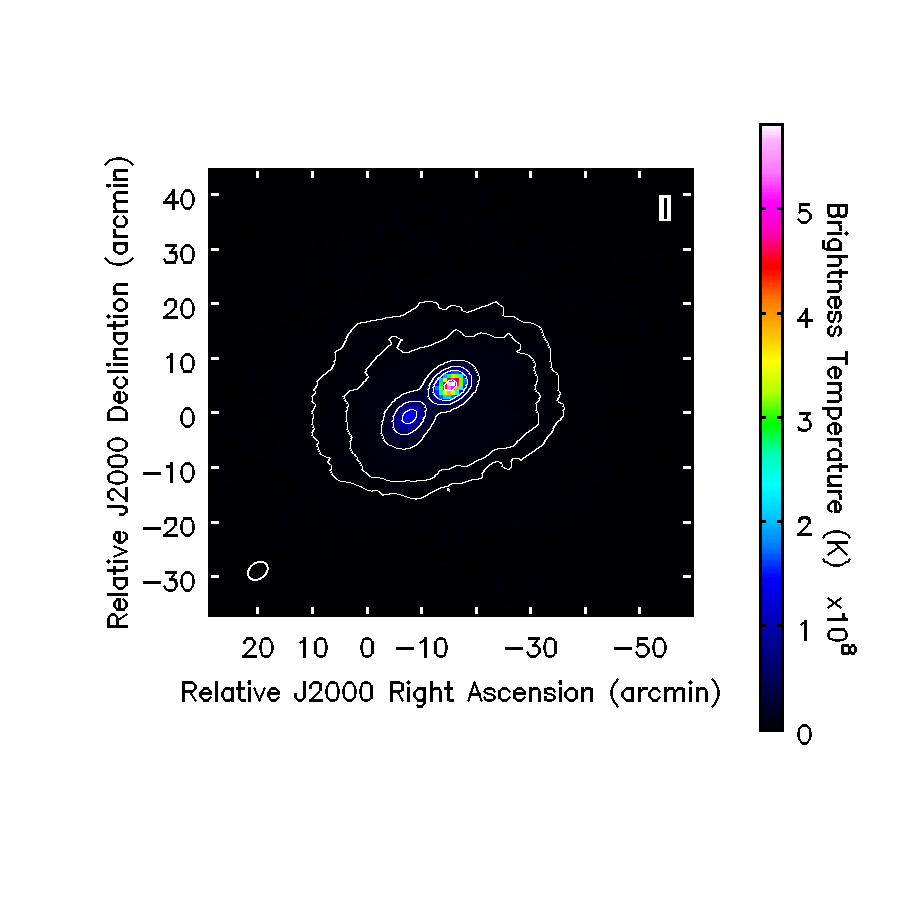}{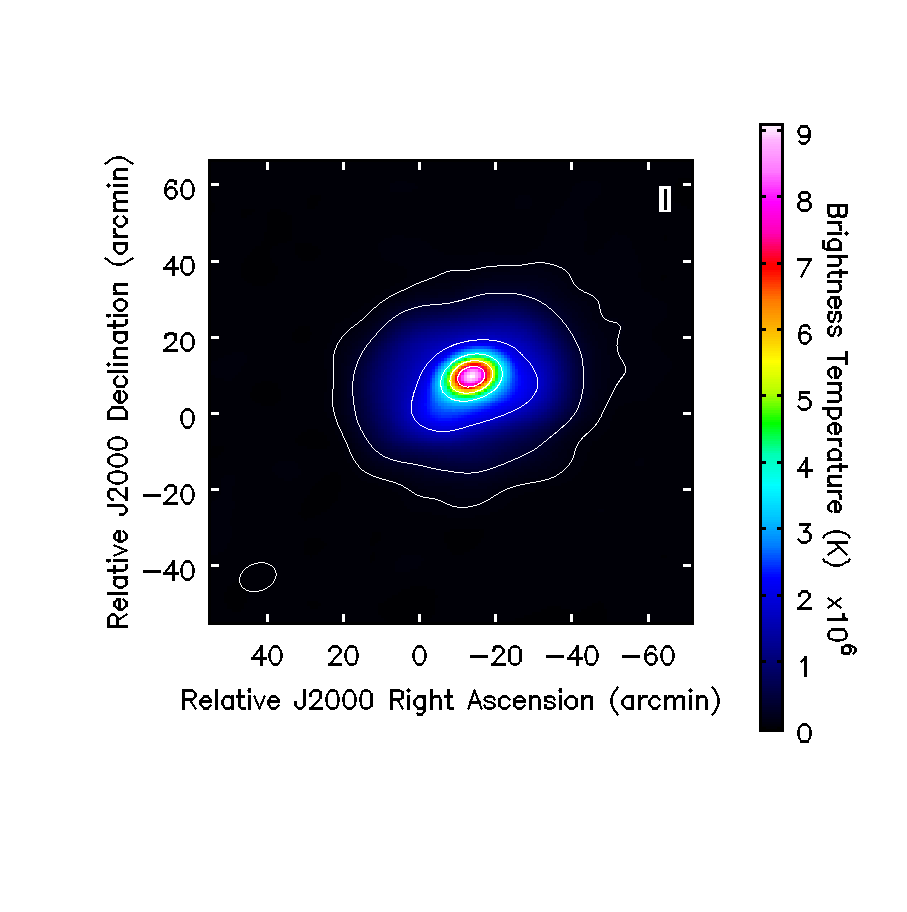}
\caption{Example Stokes I images from AIRCARS. \textit{Left upper panel} (1 of Table \ref{tab:image_properties}): Solar image during a type II burst, $T_B$~ Contours: 0.0002, 0.0003, 0.001, 0.003, 0.01, 0.03, 0.1, 0.3, 0.8, 0.9 of $10^9$ K. \textit{Right upper panel} (2 of Table \ref{tab:image_properties}): Solar image 1 minute before a type III burst, $T_B$ Contours: 0.02, 0.06, 0.2, 0.4, 0.6, 0.8, 0.9 of $1.7 \times 10^6$ K. \textit{Left middle panel} (3 of Table \ref{tab:image_properties}): Quiet sun (no sunspot visible on optical disc), $T_B$~ Contours: 0.03, 0.09, 0.4, 0.7, 0.8 of $4.0 \times 10^5$ K. \textit{Right middle panel} (4 of Table \ref{tab:image_properties}): During a type III burst, $T_B$~ Contours: 0.005, 0.008, 0.02, 0.08, 0.2, 0.8 of $5.82 \times 10^8$.  \textit{Lower panel} (5 of Table \ref{tab:image_properties}): 16s after image in right upper panel, $T_B$~ Contours: 0.01, 0.04, 0.2, 0.4, 0.6, 0.8 of $9.1 \times 10^6$. }

	\label{fig:solar_images}
\end{figure*}

We have tested the performance of AIRCARS extensively with data from the MWA Phase-I under a variety of solar conditions.
AIRCARS was configured to operate in two different modes: the first optimized to push its imaging performance without regard to computational efficiency; and the second was a run on a typical dataset to produce high time resolution spectroscopic images spanning the entire time and frequency range, keeping runtime considerations in mind.
We refer to the latter as production runs.
The datasets analysed were chosen to span a large range of solar conditions, in terms of solar flux density and distribution. 
These along with properties of some example solar images obtained are summarized in Table \ref{tab:image_properties}.
The corresponding images are shown in Fig. \ref{fig:solar_images}, in the same order.
The flux calibration of these images was done using methods described in \citet{oberoi17} and \citet{mohan17}.

\begin{figure*}[ht]
\plotsixcropped{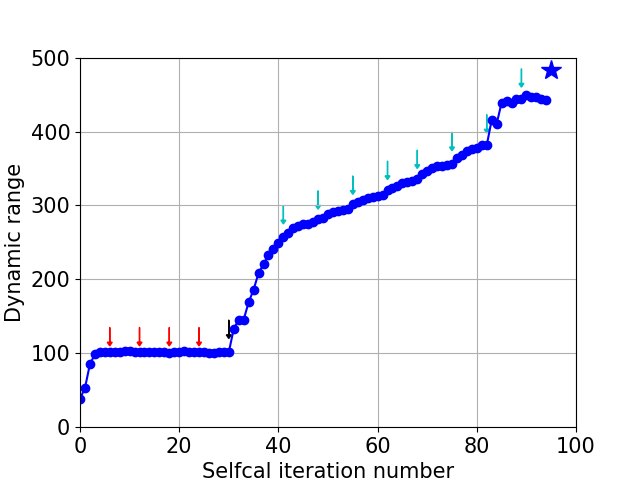}{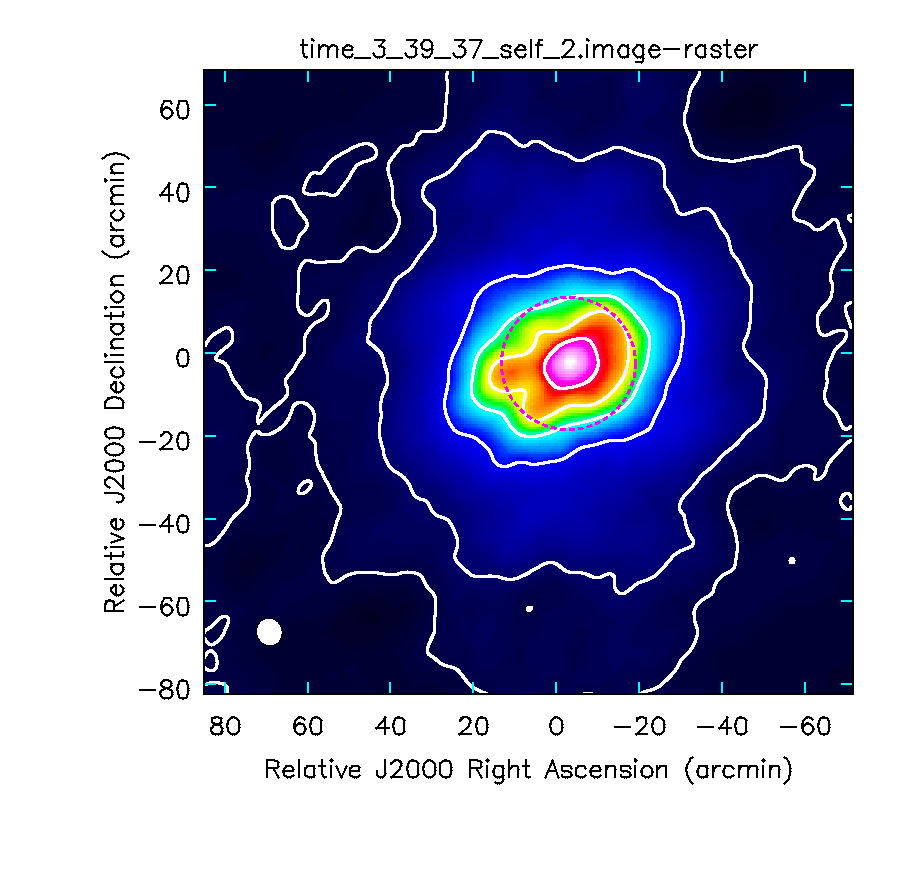}{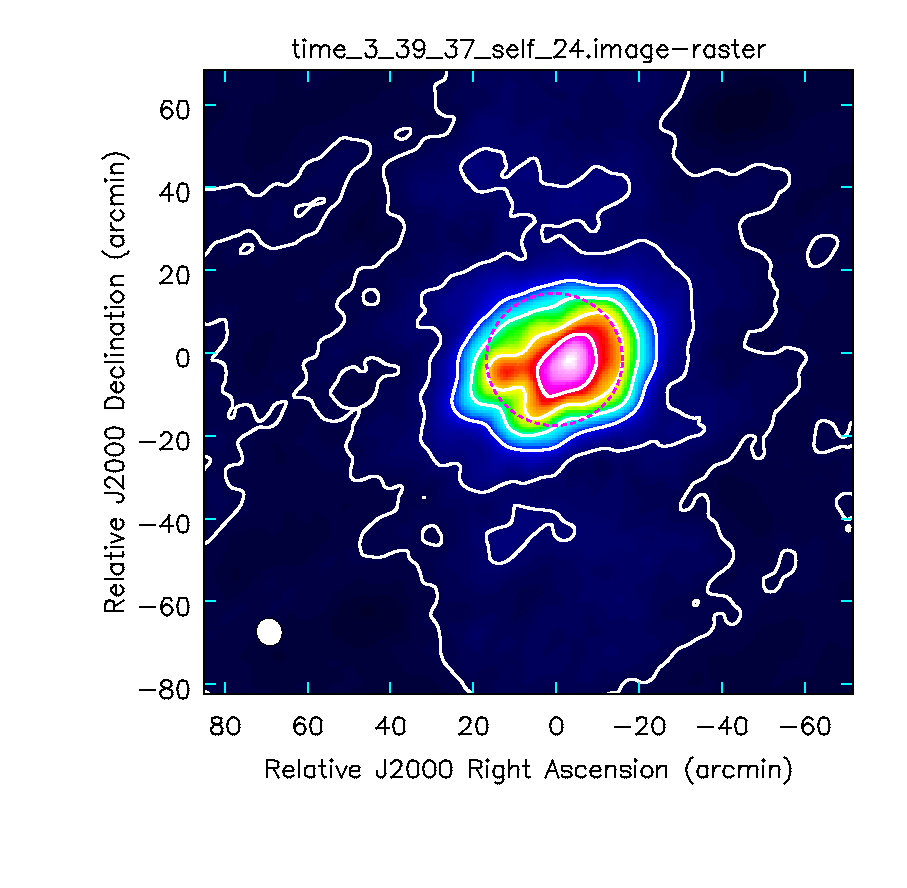}{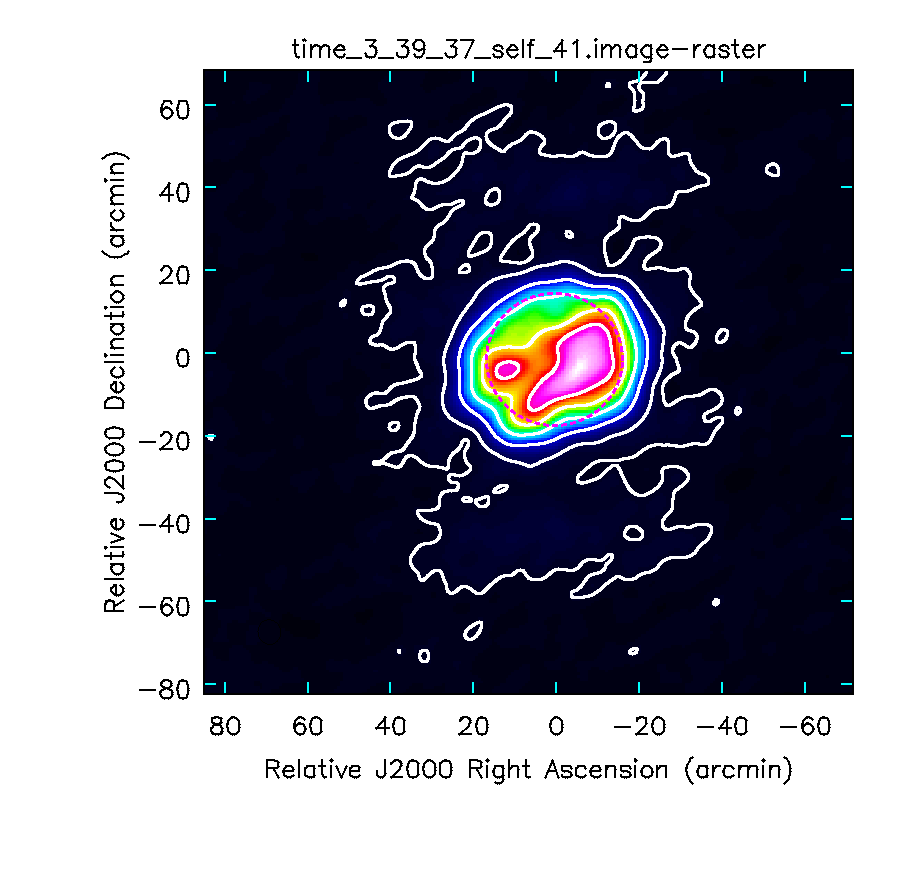}{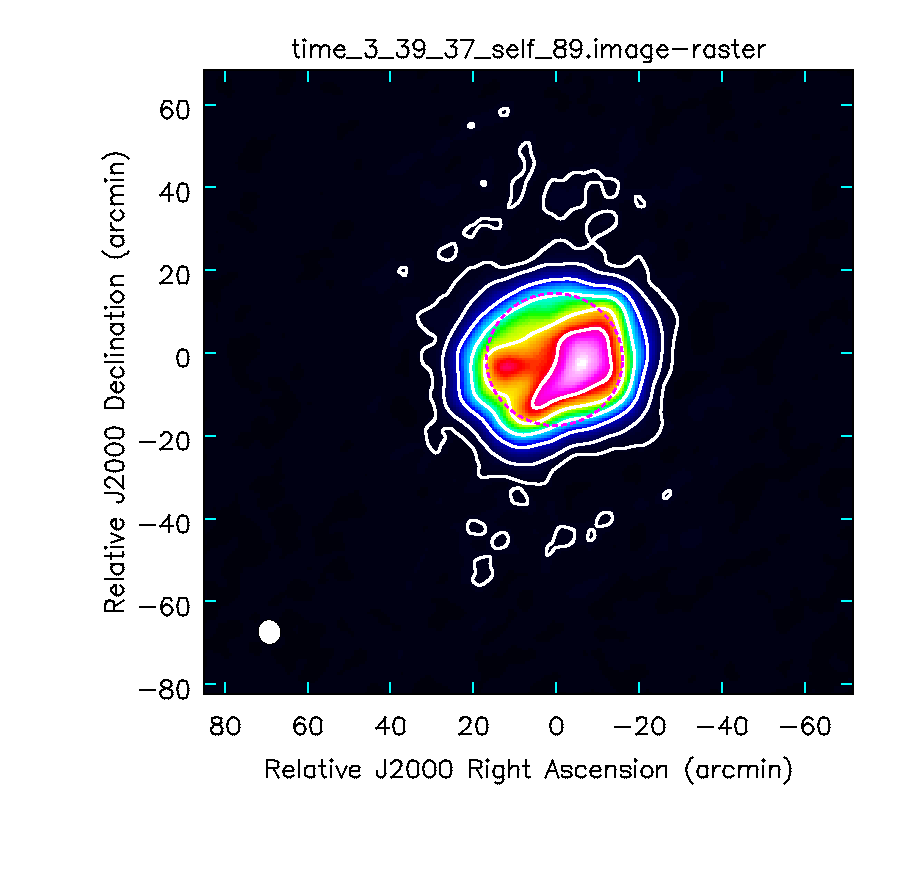}{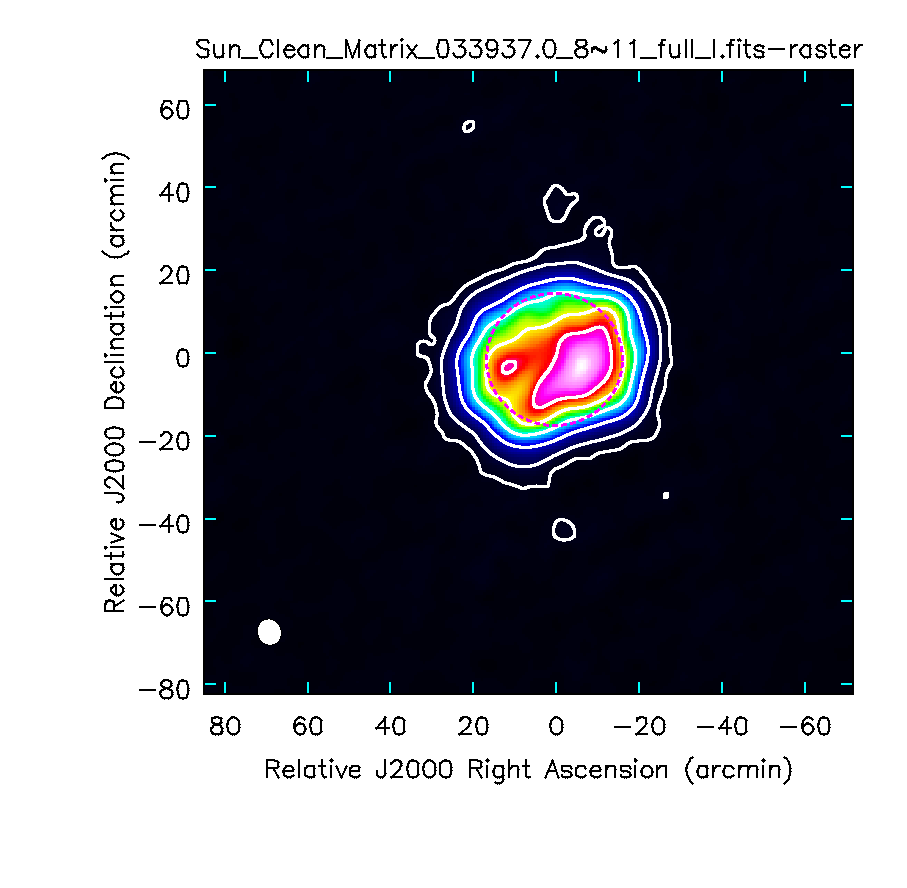}
\vspace{0.3cm}
\caption{{ \textit{Upper left panel}: Improvement of DR with self-calibration cycles. The increasing DR shows that both the $g_i$ and the source model progressively improve. The red arrows indicate the cycles where new antennas were added. The black arrow indicates the change from a phase-only to a amplitude-phase calibration. The cyan arrows indicate the cycles where $n$ was changed. The blue star shows the DR of the final deep cleaned image. 
\textit{All other panels}: Show the solar images obtained at various stages of the self-calibration process. From left to right and top to bottom, these panels show the images after 2, 24, 41, 89 self-calibration iterations respectively, and the lower right panel shows the image from the final deep clean.
The contour levels are 0.01, 0.05, 0.2, 0.4, 0.6, 0.8 times the maximum intensity in the image.
The filled white circle at the lower left corner of each figure shows the size of the restoring beam. 
The dashed magenta circle in each figure shows the optical disc of the Sun.
The improvement of image quality with the progression of self-calibration iterations is self evident.
}}
\label{fig:DR_vs_iterations}
\end{figure*}

{
In general, imaging very extended sources is significantly harder than imaging compact sources. 
We have chosen such a dataset, when no bright compact radio source was present, to illustrate the capabilities of AIRCARS.
The improvement in DR and the corresponding solar images as AIRCARS progresses through self-calibration cycles are shown in Fig. \ref{fig:DR_vs_iterations}. 
}

\begin{table*}[ht]
	\centering 
	\begin{tabular}{|C{1.5cm}|c|C{1.5cm}|C{2.0cm}|c|c|C{1.8cm}|}
		\hline
		\begin{tabular}{C{1.5cm}}Image\\number \end{tabular} & Obs. date &\begin{tabular}{C{1.5cm}}Max $T_B$\\ (MK)\end{tabular}& \begin{tabular}{C{2.0cm}}rms far\\from sun\\ (kK) \end{tabular}&\begin{tabular}{c} Dynamic\\ range \end{tabular} & \begin{tabular}{c}Central\\Frequency\\ (MHz)\end{tabular} & \begin{tabular}{c}Bandwidth\\ (kHz)\end{tabular}\\ \hline
		1 &2015/11/04 &$1050$&14& 74000 & 144.32 & 40\\ \hline
		2 & 2014/11/03 &$1.7$&2 &750 & 118.78 & 160\\ \hline
		3& 2015/12/03&$0.4$&0.4&1000&239.1 & 160\\ \hline
		4$^*$ & 2014/11/03 & $582$ & 119&4500 & 229.22 & 160 \\ \hline
		5$^*$ & 2014/11/03 & $9.1$ &7 &1290 & 112.380 & 160 \\ \hline
	\end{tabular}
	\caption{Properties of the images shown in Fig. \ref{fig:solar_images}. All the images had an integration time of 0.5\,s. The images from the production runs are marked by a $^{*}$.}
	\label{tab:image_properties}
\end{table*}


\begin{figure}
\centering
\includegraphics[trim={3.7cm 3.0cm 1.7cm 3.0cm},clip,scale=0.25]{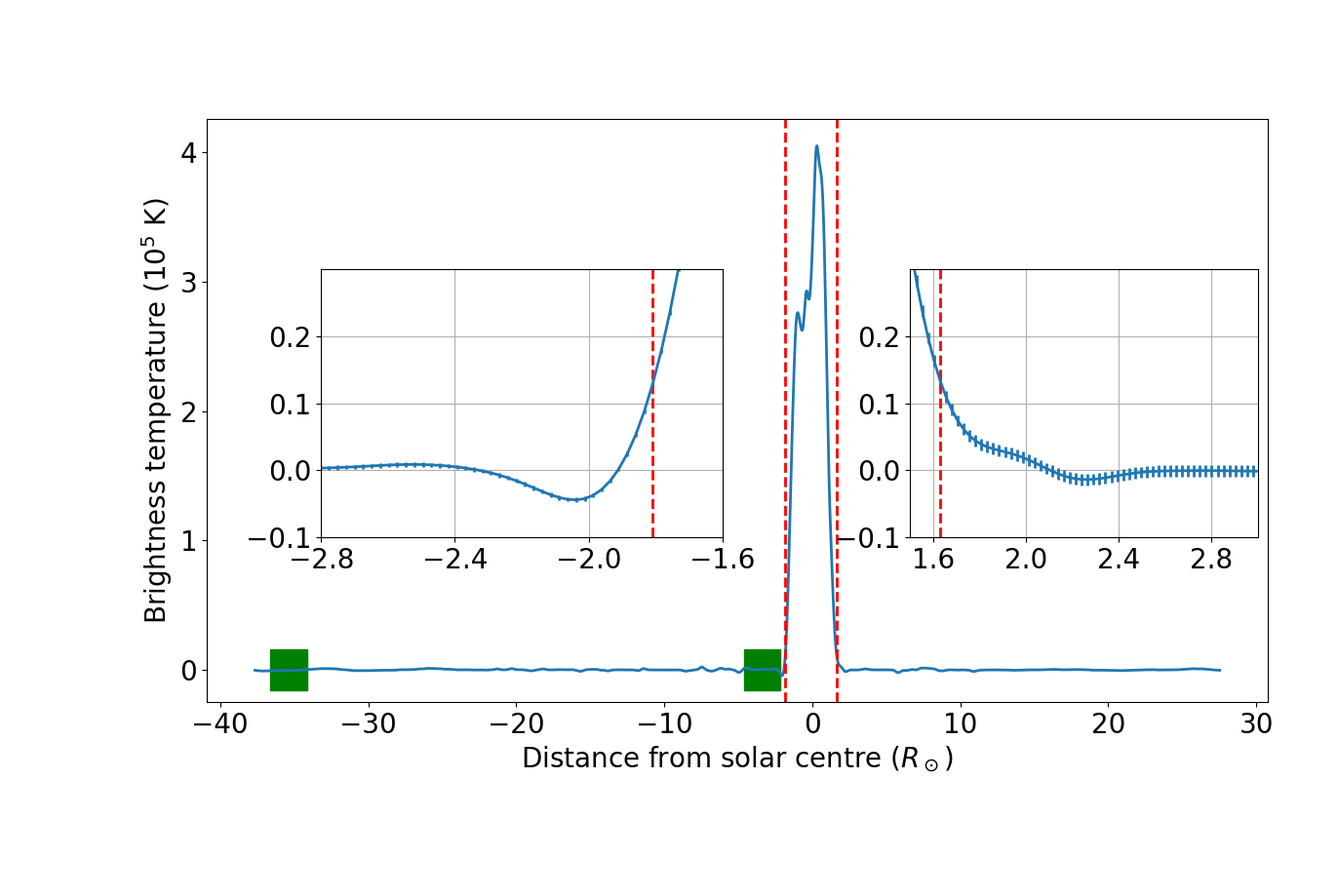}
\caption{A cut across the brightest point of the middle left panel of Fig. \ref{fig:solar_images}. The red dashed lines mark a conservative estimate of the weakest reliable flux density seen in the map, as described in the text. The rectangles shown in green show the location of the regions chosen for calculation for rms. The length of the green rectangles is same as the regions chosen for computing the rms, the height of the box is arbitrary. The actual breadth of the chosen regions are about 40$'$. The insets zoom into the regions where the red dashed lines intersect the intensity profile. The error bars shown in the left and right insets correspond to the rms calculated in the rectangles far away ($400$ K) and near ($900$ K) to the sun, respectively.}

\label{fig:1d-cut}
\end{figure}

A cut across the solar radio image in the middle left panel of Fig. \ref{fig:solar_images} is shown in Fig. \ref{fig:1d-cut}.
The location of the outermost contour (3\% of the peak) is shown by the vertical red dashed lines.
This contour represents the lowest flux density above which there is no noise peak, as defined below; hence it can serve as a conservative estimate of the weakest reliable emission detected.
The two insets zoom into the regions where the dashed red line intersects the solar intensity profile.
The observed angular size of the Sun in this image, as measured by the red dashed lines, is slightly greater than 3$R_{\odot}$.

The image rms far away from the source is conventionally regarded as a measure of the image noise floor; for this image it is $400$ K and is marked in the left inset.
The image noise characteristics, especially for strong sources, change with distance from the strong source; the rms observed in a region closer to the Sun is $900$ K and is shown in the right inset. 

The images shown here demonstrate the significant advances enabled by the confluence of the large-N architecture of the MWA; availability of modern signal processing hardware; and the AIRCARS implementation designed to squeeze the most benefits from the self-calibration process.
To provide a quantitative comparison of our snapshot high spectral resolution images with earlier work, we note that the some of the best published quiet sun radio images, in this frequency range, achieved rms $T_B$ fluctuations in the range 10-15 kK in regions outside the Sun and required seven hour synthesis observations with a bandwidth approaching 1\,MHz \citep{mercier2009}.

\section{Discussion}

\subsection{Self-calibration convergence}
\begin{figure*}
    \centering
    \includegraphics[scale=0.30]{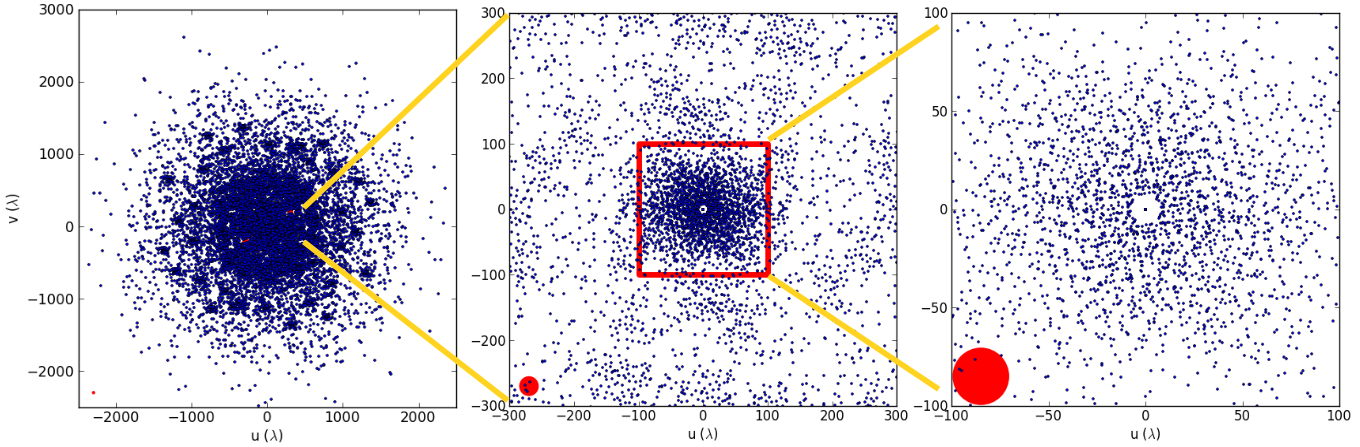}
    \caption{The snapshot uv distribution for a single frequency channel for the MWA Phase-I array is shown here. The red circle shows the uv cell size required for Nyquist sampling for a 1$^{\circ}$ FOV, the approximate angular size of the sun in our frequency range. It is evident that MWA uv sampling exceeds the Nyquist criterion over a large part of the uv plane.}
    \label{fig:uv_distribution}
\end{figure*}

Self-calibration is a highly non-linear process and its convergence to the true value is not always assured \citep{cornwell99}.
On the other hand, for an automated unsupervised analysis pipeline to be useful, the ability to produce reliable images across a wide range of solar conditions is an essential attribute. 
In our experience with the MWA data of making more than 10$^5$ images, we are yet to find an instance where this process does not converge to a high fidelity image.
We believe that it is the confluence of the following factors which places us in a regime where convergence to the global minimum seems to be assured:
\begin{enumerate}
    \item The high flux density of the Sun implies that even in the smallest time-frequency slices considered, the SNR of the signal remains sufficiently high.
    \item Even though the MWA tiles have rather large FOVs, the Sun being much brighter than other sources effectively reduces the imaging problem to that of imaging a small FOV.
    \item The array architecture of the MWA offers a number of advantages. The large number antennas ensure that the problem of solving for $g_i$ is always highly over constrained. The pseudo-random arrangement of this large number of antennas leads to a PSF with very low side-lobes.{ Together, the large number of antennas arranged in a pseudo-random maner over a small area ensure that the uv plane is densely sampled}. The angular dimensions of the Sun are such that, the sampling density in the uv plane approaches or even exceeds the Nyquist criterion over a significant part of the uv-plane (Fig. \ref{fig:uv_distribution}).
    \item Our iterative approach of building up the complexity of the model from a very simple, almost unresolved Sun to pushing it to the limits of the resolution of the array, ensures that we always start with a well conditioned prior.
    \item Conservative choices made during the AIRCARS implementation while incorporating new features into the source model minimizes the possibility of spurious structures creeping into it (Sec \ref{sec:conservative}).
\end{enumerate}

\subsection{Role of the morphology of emission and the ionosphere}
A careful examination of the images in Fig. \ref{fig:solar_images} and the information presented in Table \ref{tab:image_properties} shows that the presence of a dominant compact source correlates strongly with our ability to achieve high DR.
The noise floor achieved in all of these images is always much higher than the expectations for thermal noise, implying that one or more extraneous factors are limiting the DR achieved. 
One possible scenario is that the precision with which we can determine the $g_i$ is closely tied to the angular extent of the dominant emission on the solar disc. 

As the self-calibration iterations proceed, the estimates of $g_i$ and the sky model are expected to progressively improve. 
As mentioned in Section \ref{sec:self_calib} and \ref{sec:dense_core}, we use a DR based criterion to decide if the self-calibration process has converged.
We define a quantity called residual phase to be the difference between the phases of the $g_i$ from the penultimate and last self-calibration iterations. 
We use it as a proxy for the precision with which $g_i$ has been estimated.
For perfect calibration, the residual phase should be exactly zero.

\begin{figure*}
	\plottwo{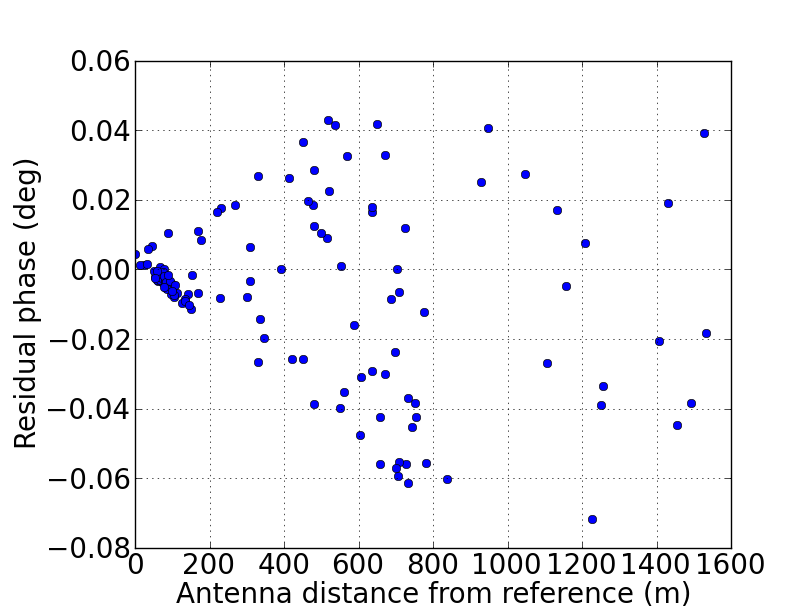}{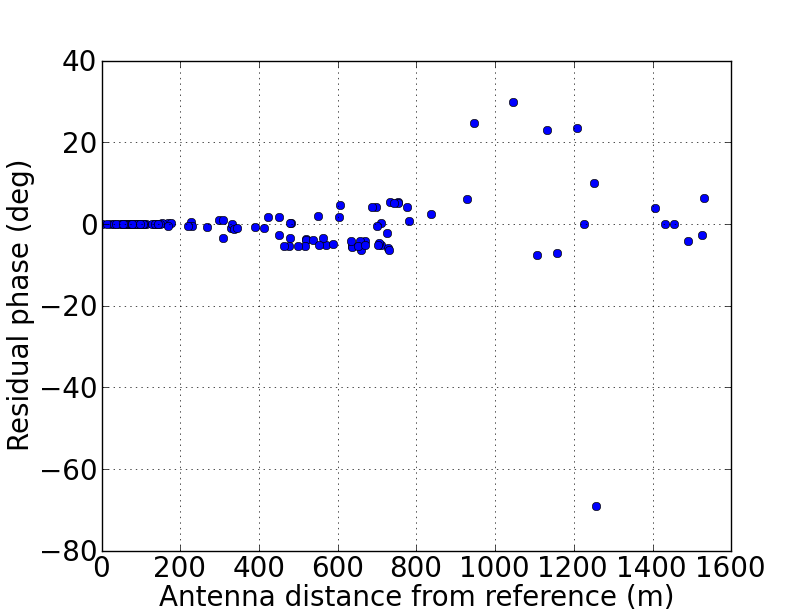}
	\caption{Antenna residual phases as a function of distance from the reference antenna. The left panel shows the residual phase determined during a dominant compact source arising from a type II burst. The right panel shows the residual phases for a case where the emission is much more uniform across the solar disc.}
	\label{fig:residual_phase}
\end{figure*}

Figure \ref{fig:residual_phase} shows a plot of residual phases for each of the antennas as their distance from the reference antenna changes, for a case when a dominant compact source was present (left panel, corresponding to top left panel in Fig. \ref{fig:solar_images}) and another one when the solar disc was much more uniform in brightness (right panel, corresponding to top right panel in Fig.  \ref{fig:solar_images}).
The figure shows that for the case of a dominant compact source, as the DR tends to saturate, the incremental changes in $g_i$ in the last two iterations are also very small ($\text{-0.08}^{\circ}$ to 0.06$^{\circ}$). 
On the other hand, in the case of a more uniform solar disc, though the DR has saturated, the $g_i$ still show significant variation from one iteration to the next (-70$^{\circ}$ to 40$^{\circ}$). 
This suggests that in the latter case there are multiple choices for $g_i$ which can differ significantly amongst themselves, but still lead to similar imaging quality.

One can envisage a hyper-surface of $\phi$ for Equation \ref{eq:selfcal_minimizer}, with the antenna gains $g_i$ and the image components as its axes. 
{In the case of a dominant compact source, we find that as the self-calibration process progresses closer to convergence, the rate of increase of DR become increasingly shallow and the observed variation in the values of $g_i$ also become very small.
Such behavior is very typical of gradient descent like minimization algorithms as they approach the true minimum.
Qualitatively, this suggests that under such circumstances, $\phi$  has a deep minimum in the vicinity of the point $\left(g_1,g_2,..,g_N, I^M_1,I^M_2,..., I^M_P \right)$, where $N$ is the total number of antennas, $I^M_i$ is the value of the $i^{th}$ component of the sky model and $P$ is the total number of the components in the model sky. 
The distance from the true minimum can be expected to be of the order of the variations in the $g_i$.
}

On the other hand, when the solar intensity distribution is much more extended and uniform, the observed behavior of a much lower DR with large residual phases, is suggestive of the $\left(g_1,g_2,..,g_N, I^M_1,I^M_2,..., I^M_P \right)$ vector lying in a shallower region of $\phi$ with multiple local minima of similar depths in its vicinity.
The observed DR and the image models do not show significant changes even though the residual phase does. This suggests that the $\phi$ surface is such that its projection on the hyper-surface spanned only by the image components has only one minima near $\left(I^M_1,I^M_2,..., I^M_P \right)$, while its projection on hyper-surface along the $g_i$ axes seems to have multiple minima of similar depths. 

An inadequacy of the model to represent the true nature of the data can lead to such behavior. 
The direction independent nature of our calibration is a known inadequacy of our model.
By assigning one number to each $g_i$ phase, AIRCARS assigns a single value for phase to the whole solar surface as seen by a given antenna, completely ignoring the existence of the ionosphere.
The presence of structures in the ionosphere on angular scales of the size of the radio solar disc of sufficient strength to have a significant detrimental impact on interferometric imaging, and the need to account for them during the imaging process, is now well established \citep[e.g.][etc.]{intema09,loi2015a,loi2015b,mevius16,jordan17}.
These studies are usually based on night time observations; the day time ionosphere can reasonably be expected to have more and varied structures.

The LOS to the solar disc samples a region of about 7 km in the ionosphere, assuming that the bulk of the electron density is concentrated in a thin layer at a height of 400 km and the size of the solar disc is about $1^{\circ}$.
Past ionospheric studies using night time data shows that ionospheric structures can have length scales ranging from 3.5 - 31 km \citep[e.g.][]{mevius16, jordan17}.
So it is expected that the ionospheric electron density varies across the area sampled by the LOS to the solar disc.

\begin{figure*}
    \centering
    \includegraphics[scale=0.35]{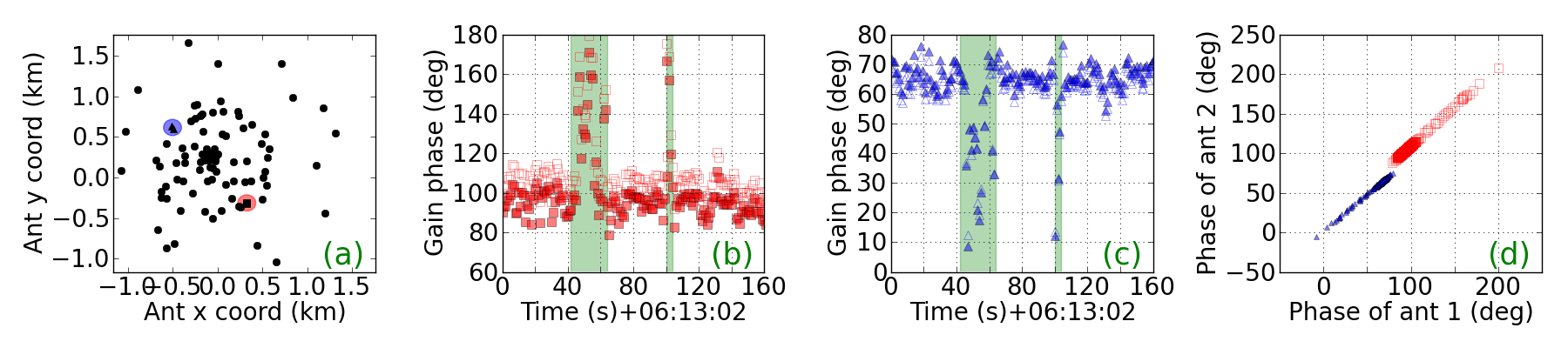}
    \caption{Panel \textit{a} shows the location of the antenna pairs. Antennas marked by triangle inside the blue patch is a chosen pair and antennas marked by square inside the red patch forms the other pair. In panel \textit{b} and \textit{c} the actual phase timeseries of the antenna pairs are shown. The phases of the two elements of a pair are denoted by filled and hollow symbols. The symbols (square or triangle) and colour (red or blue) are same for each pair in all the panels. The green shaded areas mark the times where we see large phase variations. In panel \textit{d}, the phase of one antenna of a pair is plotted against the phase of the other antenna as a function of time.} 
    \label{fig:phase_variation_with_time}
\end{figure*}

In Figure \ref {fig:phase_variation_with_time}, phase variations of two antenna pairs with time are shown.
The two pairs are located in different parts of the array and are chosen such that the distance between the two antennas of a pair is very small.
The phase variations are shown in panels \textit{b} and \textit{c}.
The two circular patches in panel \textit{a} show the locations of the antenna pairs in the array.
The pair in the red patch is indicated by square markers and that in the blue patch by triangular markers.
The phases determined as a function of time by AIRCARS for antennas in the red patch are shown in panel \textit{b} using filled and hollow red squares.
Similarly those for the blue patch are shown in panel \textit{c} using filled and hollow blue triangles.
The same data are shown in panel \textit{d} (red squares and blue triangles correspond to the antenna pairs shown in panels \textit{b} and \textit{c}, respectively) in a different representation to highlight their highly correlated nature. 

Noise fluctuations in antennas forming a pair must be independent.
As evidenced from their highly correlated nature (panel d), the about $\pm10^{\circ}$ phase variations seen outside the shaded regions, which at first glance might seem like noise, are in fact reliable measurements of changes in $g_i$.
This implies that these phase fluctuations must have a common physical origin. 

The observed behaviour of these phase variations are consistent with expectations of their having an ionospheric origin.
A phase change of $10^{\circ}$ at 108 MHz corresponds to a differential change in the ionospheric electron column density by 2 mTECU. To the best of our knowledge, such rapid but weak variations in ionospheric TEC have not been reported earlier.

In all the cases which we have examined so far, a sudden change in $g_i$ phase (highlighted in panels b and c) is always accompanied by a sudden change in the solar emission morphology (both in terms of the intensity and its distribution).
This is explained by the facts that the $g_i$ phase essentially corresponds to a weighted average over the entire ionospheric screen lit up by the Sun; and a sudden change in the emission morphology leads to a corresponding change in the weighting function and hence also the observed value of the $g_i$ phase.
This also emphasizes that the requirements of isoplanicity for very high dynamic range solar imaging are much more stringent than those in conventional radio imaging.

\subsection{AIRCARS and sub-second phase variations}
The rapid evolution of the $g_i$ phases imply the need for an independent  self-calibration even on sub-second time scales. 
The nominal value of tolerance within which the variation of $g_i$ phase will not lead to a discernible negative impact on imaging depends on multiple aspects.
These include the nature of ionospheric conditions prevailing during the observations; the nature and dynamics of solar emission morphology; and the imaging DR requirements imposed by a particular science objective. It is hence not feasible to determine the interval over which to compute independent self-calibration solutions {\em a priori}. 

Some experimentation is helpful, keeping in mind that a smaller $\tau_{cal}$ will lead to better imaging performance at the cost of a larger computational burden. As a fail-safe feature to protect against issues due to sudden changes in observed $g_{i,prop}$, AIRCARS checks for a sudden drop in the DR obtained when using a solution from a near by time slice. The maximum value of the permitted drop is a user defined parameter. If a drop larger than this threshold is observed, a fresh self-calibration solution is obtained for the time stamp in question.

\section{Conclusion}
 In this paper, we present a new algorithm optimised for imaging the Sun at low radio frequencies using arrays with compact cores.
 We also present AIRCARS, an end-to-end imaging pipeline, where we have implemented this algorithm. 
 AIRCARS has been tested so far using data from the MWA Phase-I array under a variety of solar conditions.
 We demonstrate the large improvements in imaging DR brought about by the use of AIRCARS.
 The highest DR image (DR $\sim$ 75000) obtained by using AIRCARS on MWA data is also presented in this paper.
{ The novel features which enable AIRCARS to deliver such high fidelity and high DR images are:}
 \begin{enumerate}
     \item {AIRCARS uses the advantages offered by the centrally dense footprint of the array to the hilt. 
     Starting with the dense and compact array core allows it to achieve robust gain calibration while using a very simple source model, a barely resolved Sun.
     The progressive inclusion of antennas at larger distances from the core, ensures that a small number of new degrees of freedom are introduced at a manageable rate and under circumstances where the vast majority of them are already close to their optimal estimates.
     This allows AIRCARS to start with little prior information about the emission model and the antenna gains, and slowly build up complexity in the source model. 
     }
     \item {The completely unsupervised operation of AIRCARS makes it feasible to perform an independent self-calibration of every time and frequency slice of the data, and tune AIRCARS in favor of imaging performance.
     This, combined with the high SNR provided by the strong solar signal allows AIRCARS to account for the tiny gain variations over sub-second scales and capture the small changes in the emission morphology over very small fractional bandwidths.
     Neither of these are practical without a very high degree of automation in the analysis.
     }
     
     \item {Heuristics, deduced from extensive prior solar imaging experience with the MWA, are incorporated in AIRCARS. This enables it to tune the self-calibration process to the data presented to it.
     }
 \end{enumerate}
As mentioned, AIRCARS can be tuned for high DR performance or optimized for run-time to churn through large data volumes. 
With the eventual objective of making AIRCARS easy to use for the broader community, AIRCARS comes with default values of practically all calibration and imaging related parameters. These defaults have been tested by imaging large data sets and are oriented towards run time optimization. 

Though this work represents an enormous improvement in the state-of-the-art in solar radio imaging at these frequencies, the MWA data themselves are capable of even more.
The dependence of DR observed on the nature of observed solar morphology implies that something extraneous is currently limiting the obtained DR, most likely the direction dependent effects arising due to the  ionosphere.
Making further progress will require taking this into account, and an effort to do this is currently underway.

AIRCARS is available on request. By making state-of-the-art solar radio imaging with telescopes like the MWA and the SKA-Low in future, accessible to the non-specialist, we believe AIRCARS can mark a significant step forward in greater use of these very interesting and informative data in the larger solar and heliospheric physics community.

\acknowledgments 
This scientific work makes use of the Murchison Radio-astronomy Observatory, operated by the Commonwealth Scientific and Industrial Research Organisation (CSIRO). We acknowledge the Wajarri Yamatji people as the traditional owners of the Observatory site. Support for the operation of the MWA is provided by the Australian Government through the National Collaborative Research Infrastructure
Strategy (NCRIS), under a contract to Curtin University administered by Astronomy Australia Limited. We acknowledge the Pawsey Supercomputing Centre, which is supported by the Western Australian and Australian Governments.
DO acknowledges numerous enjoyable and fruitful discussions with Sanjay Bhatnagar (NRAO). SM acknowledges fruitful discussions with Huib Intema. SM also acknowledges Barnali Das for coming up with a nice name for the software.
CJL and LB acknowledge financial support from US Air Force Office of Scientific Research (AFOSR) award FA9550-14-1-0192.
IHC acknowledges financial support from Australian Research Council (ARC) grant DP180103509 and US Air Force Office of Scientific Research (AFOSR) award FA9550-18-1-0671.
We thank the developers of Python 2.7\footnote{See
https://docs.python.org/2/index.html.} and the various associated packages, especially Matplotlib\footnote{See http://matplotlib.org/.}, Astropy\footnote{See http://docs.astropy.org/en/stable/.} and NumPy\footnote{See https://docs.scipy.org/doc/.}. This research has made use of NASA's Astrophysics Data System.

\facility{The Murchison Widefield Array}
\software{CASA, Python, Numpy, Astropy, Matplotlib}
\bibliography{bibligraphy}

\end{document}